\documentclass[letter]{aa} 

\usepackage{graphicx}
\usepackage{txfonts}
\begin{document}

   \title{Resolving the stellar activity of the Mira AB binary with ALMA}

   \author{W.~H.~T.~Vlemmings\inst{1}\fnmsep\thanks{wouter.vlemmings@chalmers.se}
          \and S.~Ramstedt\inst{2} 
          \and E.~O'Gorman\inst{1}
          \and E.~M.~L.~Humphreys\inst{3}
          \and M.~Wittkowski\inst{3}
        \and A.~Baudry\inst{4},\inst{5}
      \and M.~Karovska\inst{6}}

   \institute{Department of Earth and Space Sciences, Chalmers University of Technology, Onsala Space Observatory, 439 92 Onsala, Sweden
     \and
            Department of Physics and Astronomy, Uppsala University,
            Box 516, SE-751 20 Uppsala, Sweden
       \and
           ESO Karl-Schwarzschild-Str. 2, 85748 Garching, Germany
       \and
          Universit\'e de Bordeaux 1, LAB, UMR 5804, F-33270, Floirac, France
       \and
          CNRS, LAB, UMR 5804, F-33270, Floirac, France
       \and
            Smithsonian Astrophysical Observatory, MS 70, 60 Garden Street, Cambridge, MA 02138, USA
            }

   \date{Submitted: 23-03-2015, Accepted 18-04-2015}

 \abstract
{}
 {We present the size, shape and flux densities at millimeter continuum
   wavelengths, based on ALMA science verification observations in Band 3
   ($\sim94.6$~GHz) and Band 6 ($\sim228.7$~GHz), from the binary Mira
   A (${\rm o}$~Ceti) and Mira B. }
  {The Mira AB system has been observed with ALMA at a spatial
    resolution of down to $\sim25$~mas. The extended atmosphere of
    Mira A and the wind around Mira B
    sources are resolved and we derive the size of Mira A and of the
    ionized region around Mira B. The spectral indices within
    Band 3 (between 89--100\,GHz) and between Band 3 and Band 6 are also derived.}
  {The spectral index of Mira A is found to change from 1.71$\pm$0.05
    within Band 3 to $1.54\pm0.04$ between Band 3 and 6.  The spectral
    index of Mira B is 1.3$\pm$0.2 in Band 3, in good agreement with
    measurements at longer wavelengths. However it rises to
    $1.72\pm0.11$ between the bands. For the first time the extended
    atmosphere of a star is resolved at these frequencies and for Mira
    A the diameter is $\sim3.8\times3.2$~AU in Band 3 (with brightness
    temperature $T_b\sim5300$~K) and $\sim4.0\times3.6$~AU in Band 6
    ($T_b\sim2500$~K). Additionally, a bright hotspot $\sim0.4$~AU and
    with $T_b\sim10000$~K is found on the stellar disc of Mira A. The
    size of the ionized region around the accretion disk of Mira B is
    found to be $\sim2.4$~AU. }
  {The emission around Mira B is consistent with that from a
    partially ionized wind of gravitationally bound material from Mira
    A close to the accretion disk of Mira B. The Mira A
    atmosphere does not fully match predictions, with brightness
    temperatures in Band 3 significantly higher than expected,
    potentially due to shock heating. The hotspot is likely due to
    magnetic activity and could be related to the previously observed
    X-ray flare of Mira A.}

   \titlerunning{ALMA resolves the Mira binary}
   \authorrunning{W.~H.~T.~Vlemmings et al.}

   \maketitle
%
\section{Introduction}

Mira is the closest symbiotic binary star (at 92~pc,
\citet{vanLeeuwen07}) consisting of a regularly pulsating, mass-losing
Asymptotic Giant Branch (AGB) primary, Mira A (${\rm o}$ Cet), and a
companion, Mira B (VZ Cet), believed to be a white dwarf. Resolved UV
observations of the two components were first published in
\citet{karoetal97} and revealed material flowing from Mira A to Mira B. 

A soft X-ray outburst was detected from Mira A in 2003
\citep{karoetal05}. Since then, the system has been closely monitored
at different wavelengths to further understand the cause and effects
of the X-ray outburst. The OH maser emission is also varying and the
most recently observed OH maser flare occured in 2009
\citep{etoketal10}. Although the OH flaring can be correlated with
variations in the optical lightcurve and the UV line emission occuring
on the same time scales, a consistent model explaining the transient
phenomena detected by different probes is still missing.

The submm emission from the Mira AB system around the
CO($J$=3$\rightarrow$2) line at 345\,GHz was observed with ALMA in
Cycle 1 \citep{ramsetal14}. The CO maps show that the circumstellar
gas has been shaped by several different dynamical processes during
the evolution of the star. In the continuum emission centered on
338\,GHz, the binary pair is marginally resolved. In this paper, we
present new ALMA continuum observations of the system taken as part of
the ALMA long-baseline science verification campaign (ALMA
partnership, 2015).

\begin{figure*}
\sidecaption
\includegraphics[width=12.5cm]{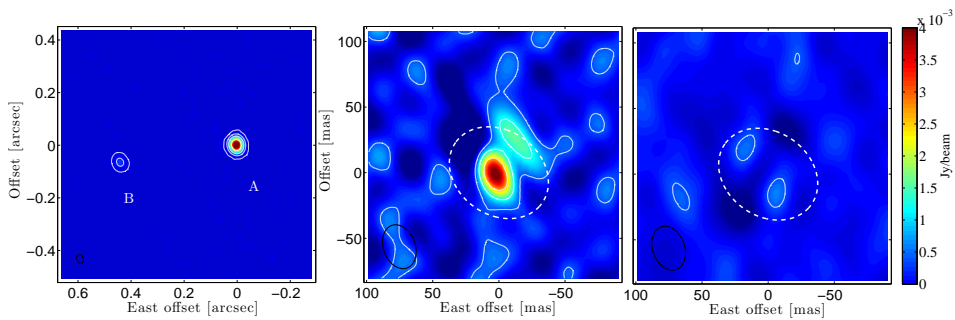} 
\caption{Continuum image of the Mira AB binary at Band 6 {\it (left)}. Residuals in Band 6 at the location of
 Mira A after subtracting the best fit disc model {\it (middle)} and
 the disc model including a compact gaussian hotspot {\it (right)}
 from the visibilities. The dashed ellipse indicates the size of the
 fitted stellar disc. Contours in all three plots are drawn at $3, 30, 90, 270,
  810\sigma$ with  $\sigma=140~\mu$Jy~beam$^{-1}$. The beam size is shown
  in black in the bottom left corner of all three figures.}
\label{fig:Mira2}
\end{figure*}

\section{Observations and analysis}
\label{obs}
Mira was observed on 17 and 25 October 2014 (Band 3) and 29 October
and 1 November 2014 (Band 6). Here we present results from the
dual polarization continuum spectral windows. In Band 3, three continuum spectral
windows were observed with 128 channels over 2~GHz bandwidth centered
on 89.176~GHz, 91.234~GHz and 99.192~GHz. In Band 6, one continuum
2~GHz spectral window with 128 channels was centered at
228.558~GHz. 
We used the data products provided by the ALMA observatory except that
the two epochs for each observing band were treated separately as
  the sources are strongly detected in each epoch, allowing for the
  study of structural and flux changes. Self-calibration was
performed on the continuum spectral windows that were averaged from
128 to 8 channels across the 2~GHz bandwidth. Flux density calibration
was performed using a number of different quasars for the two bands
and epochs. The calibrators are listed in Appendix~\ref{flux}. Based
on the uncertainty of the catalog values due to quasar variability and
the flux densities measured for the phase and bandpass calibrators in
the different data sets, we conservatively estimate an {\it absolute}
flux density calibration uncertainty of $\sim5\%$ in both observing
bands.

Most of the data analysis was performed directly on the $uv$-data in order to avoid image deconvolution artifacts, but
images were also created using the CASA task {\it clean} with Briggs
$0.5$ weighing of the visibilities. 
This resulted in a beam of $72\times64$~mas and a position angle of
$78.8^\circ$ on October 17 and $68\times57$~mas at a position angle of
$73.3^\circ$ on October 25 for Band 3. For Band 6 both days reached
the same resolution of $34\times24$~mas with a position angle of
$22.4^\circ$ on October 29 and $16.0^\circ$ on November 1. The rms
sensitivity was $\sim40~\mu$Jy~beam$^{-1}$ in both epochs of the Band
3 and $\sim130~\mu$Jy~beam$^{-1}$ in both epochs of the Band 6
observations.  The images in Band 3, with $>10\%$ fractional bandwidth
coverage, were made using a the multi-frequency synthesis option in
{\it clean} with two Taylor coefficients to model the frequency
dependence. This allowed us to determine the spectral index at the
$\sim94.2$~GHz representative frequency.

To analyze the $uv$-data directly, we used the UVMULTIFIT code
\citep{Marti14}. We find that the observations of October 25 in Band 3
are likely affected by a marginally resolved quasar flux calibrator,
producing systematically smaller sizes, and do not consider this epoch
in our analysis.


\section{Results}
\label{res}

We present a representative image at the two frequencies in
Fig.~\ref{fig:Mira2}~(left, band 6) and
  Fig.~\ref{fig:Mira}~(left, band 3). The relative astrometry of the observations
is discussed in Appendix~\ref{relastrom}.  In order to determine the
size and shape of both Mira A and Mira B, we performed
$uv$-fitting. For both sources we attempted fits to unresolved delta
functions, elliptical and circular Gaussians and elliptical and
circular discs. The best fit parameters for Mira A and B at each epoch
and observing band are presented in Table~\ref{uvfit}. As indicated in
Fig.~\ref{fig:Mira}~(right), the emission from Mira B is clearly
extended and the best fits were obtained using a circular Gaussian
with a spectral index of $\alpha=1.3$. Within the fitted errors and
estimated {\it absolute} flux density uncertainties, the flux density of Mira B is
roughly constant at the two epochs in both bands, although in Band 3
there might be a hint of variability at the $5-10\%$ level. Between
Band 3 and Band 6, the spectral index is $1.72\pm0.11$, taking into
account all flux density uncertainties. 
Ignoring the observations of October 25, the size is consistent between the different
frequencies with a full-width half-maximum (fwhm) of $\approx26$~mas,
which corresponds to $\sim2.4$~AU. 

For Mira A, the best fits to the Band 3 data were produced with an
elliptical disc with $\alpha\approx1.7$. The flux density is nearly
constant during both days at each observing frequency. The spectral
index between Band 3 and Band 6 however, is shallower with
$\alpha=1.54\pm0.04$. The disc is clearly elongated, with a major axis
of $41.8$~mas ($3.8$~AU), axis ratio of $\sim1.2$ and position angle
$\sim54^\circ$.  However, in Band 6, we found as indicated in
Fig.~\ref{fig:Mira2}, a significant component remaining after
subtracting the best fit disc model. This component, consistently at
both epochs, could be fit by an additional compact gaussian offset by
$\sim3$~mas from the disc center. Fitting both disc and gaussian
simultaneously signicantly reduced the residuals in the image. The
size of the disc is larger at Band 6 compared to Band 3, with an
average size of $43.32$~mas (3.99~AU) and an axis ratio of $\sim1.1$.

\begin{figure*}
\sidecaption
\includegraphics[width=12.5cm]{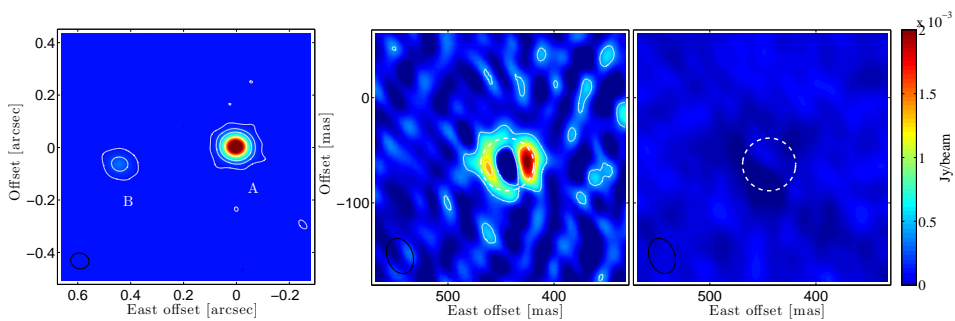} 
\caption{Continuum image of the Mira AB binary at Band 3 {\it (left)}. Residuals are shown for
  Band 6 at the location of Mira B after subtracting the best fit
  delta function {\it (middle)} and extended Gaussian {\it (right)}. The dashed circle indicates the fwhm of the
  fitted Gaussian. Contours in all three plots are drawn at $3, 30, 90, 270,
  810\sigma$ with $\sigma=40~\mu$Jy~beam$^{-1}$ in Band 3 and
  $\sigma=140~\mu$Jy~beam$^{-1}$ in Band 6. The beam size is shown
  in the bottom left of all three figures.}
\label{fig:Mira}
\end{figure*}

\begin{table*}
\caption{$uv$-fitting results}
\begin{center}
\begin{tabular}{lccccccc}
\hline \hline
epoch & $\nu$ & shape & $S_{\nu}$ & major axis / fwhm & axis ratio & position angle & spectral index\\
         & [GHz]  &       &   [mJy]       & [mas]      & major/minor & [$^{\circ}$] & \\
\hline \hline
Mira A & & & & & & & \\
\hline
17 Oct 2014  &  94.2 &  Disc & 35.03$\pm$0.04  &  41.8$\pm$0.4 &  1.20$\pm$0.01 &  54$\pm$2 & 1.73$\pm$0.09 \\
25 Oct 2014$^a$  &  94.2 &  Disc & 34.52$\pm$0.04  &  38.6$\pm$0.4 &  1.28$\pm$0.01 &  54$\pm$1 & 1.70$\pm$0.04 \\
29 Oct 2014  &  228.67 &  Disc & 137.8$\pm$0.2 &  43.28$\pm$0.07 &  1.13$\pm$0.02 &  51.0$\pm$0.5 & \ldots \\
     &  228.67 &  Gaussian  & 10.13$\pm$0.07 & 4.6$\pm$0.5 & 1.0 &  \ldots & \ldots \\ 
01 Nov 2014    &  228.67 &  Disc & 140.0$\pm$0.2 &  43.36$\pm$0.06 &  1.12$\pm$0.02 &  50.8$\pm$0.6 & \ldots\\
     &  228.67 &  Gaussian & 8.98$\pm$0.07 & 4.7$\pm$0.5 & 1.0 &  \ldots & \ldots\\
\hline 
Mira B & & & & & & & \\
\hline
17 Oct 2014      &  89.06 &  Gaussian & 2.50$\pm$0.04  &  24$\pm$2 &  1.0 & \ldots & 1.3$\pm$0.2 \\
25 Oct 2014$^a$      &  89.06 &  Gaussian & 2.25$\pm$0.04  &  18$\pm$2 &  1.0 & \ldots & 1.3$\pm$0.3 \\
29 Oct 2014  &  228.67 &  Gaussian & 12.16$\pm$0.09  &  25.7$\pm$0.3 &  1.0 & \ldots & \ldots \\
01 Nov 2014  &  228.67 &  Gaussian & 11.98$\pm$0.09  &  26.2$\pm$0.3 &  1.0 & \ldots & \ldots\\
\hline \hline
\multicolumn{8}{}{}{\footnotesize $^a$ The size of both Mira A and B are likely
underestimated at the second epoch due to a marginally extended flux
calibrator J2258-2758} \\
\end{tabular}
\end{center}
\label{uvfit}
\end{table*}%

\section{Discussion}
\label{dis}

\subsection{Mira A}
\subsubsection{Size and shape}
Mira A has long been known to have an asymmetric stellar disc
\citep[e.g.][]{Karovska91, Haniff92, Wilson92, Quirrenbach92}. The
position angle and axis ratio however, has been shown to be quite
variable across the stellar phase.  Most observations taken around
maximum phase are found to have a position angle of the major axis
between $\sim105-160^\circ$.
Closer to the phase of the ALMA observations (phase $\sim0.36$) the
observations of \citet{Karovska91} (at phase $0.43$) find a position
angle of $\sim30^\circ$. The elongation itself varies with wavelength
and phase from none to $\sim20\%$, consistent with our
observations. The wavelength dependence of the elongation is likely
due to opacity effects. The apparent variation of position angle with
stellar phase could indicate that stable non-radial pulsations shape
the extended atmosphere.

The diameter of Mira A also varies with phase by about $10\%$ and was
measured to be $\sim32$~mas at $2~\mu$m wavelength
\citep{Woodruff04}. This corresponds to a Rosseland radius of
$\sim1.5$~AU. The radius derived from the major axis we measure at Band 3
would thus corresponds to $\sim1.3$~R$_*$ while at Band 6 it
corresponds to $\sim1.35$~R$_*$. 
Previous observations at 43~GHz resolved the radio photosphere of Mira
A and derived a uniform disc diameter of $\sim52$~mas, corresponding
to $\sim1.6$~R$_*$ \citet{reidment07}. The fact that the Band 6
observations appear to probe a larger radius than the Band 3
observations likely requires different opacity sources throughout the
atmosphere.

\subsubsection{Emission process}
\begin{figure}
 \centering
 \includegraphics[width=8.5cm]{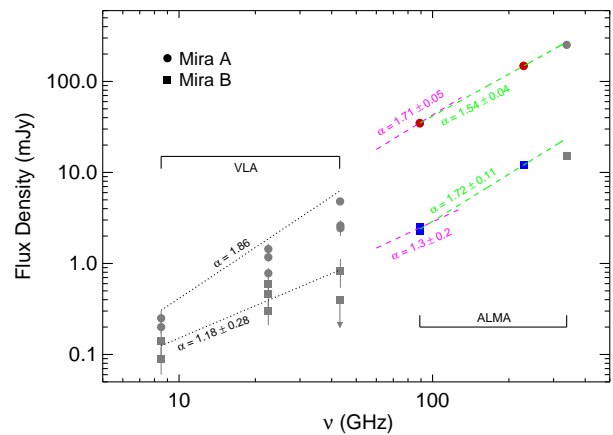} 
 \caption{The Mira A and B spectral energy distribution from the radio
   to the sub-millimeter frequency range as probed by ALMA and the
   VLA. The round (red) symbols denote Mira A and the squares (blue)
   Mira B. The grey symbols are literature referred in the text.
 The errors on the VLA data points only contain
   the formal flux density uncertainties, while the ALMA error bars also
   contain the absolute uncertainties ($5\%,$ and $10\%$ for Band 3/6,
   and 7 respectively). The dotted lines indicate the
   \citet{reidment97} model for Mira A and the fit to the VLA data for
   Mira B. The dashed line at the Band 3 points is the spectral index
   derived within the Band and the green dashed line is the fit to the
   ALMA Band 3 and 6 data.}
 \label{fig:sed}
\end{figure}

\citet{reidment97} analyze different centimeter-wavelength continuum
emission processes and conclude that the detected emission at
8.5--22\,GHz originates in a radio photosphere reaching unity optical
depth at around $\sim$2\,R$_{\star}$, predicting
spectral index of $\alpha$=1.86 at centimeter
wavelengths. \citet{mattkaro06} published resolved images of the Mira
system at 8.5--43.3\,GHz. The system was monitored with the VLA over
$\approx$80\% of the pulsational cycle and showed flux density variability
$\lesssim$30\% below 22~GHz. The lower frequency data is consistent
with the model from \citet{reidment97}. At 43\,GHz the flux density
appears to show larger variability based on the observations by
\citet{reidment07} who found a flux density almost a factor of two
higher than that reported in \citet{mattkaro06}. The VLA values, the
ALMA values from this work and the flux density at 345~GHz
\citep{ramsetal14} are presented in Fig.~\ref{fig:sed}. The most
striking result is that the spectral index between the ALMA Band 3 and
Band 6 observations ($\alpha=1.54\pm0.04$) appears significantly more
shallow than that within Band 3 itself ($\alpha=1.70\pm0.05$). The
shallower slope agrees with the observations in ALMA Band 7 though the
steeper slope would be required to fit the radio observations. Both
spectral indexes are not quite in agreement with the model prediction
by \citet{reidment97}, indicating that at millimeter wavelengths the
sources of opacity change. Our observations yield significantly higher
brightness temperature than expected from the VLA observations. For a uniform disc we find
$T_b\sim5300$~K at Band 3 ($\sim1.3$~R$_*$) and $T_b\sim2500$~K at
Band 6 ($\sim1.35$~R$_*$), compared to the $T_b\sim1650$~K at 43~GHz
($\sim1.6$~R$_*$). Only part of this difference could be explained by
variability and especially the higher temperatures in Band 3 might
require shock heating of the atmosphere close to the stellar surface.

\subsubsection{Stellar Activity}
As shown in Fig.\ref{fig:Mira2}, the Band 6 data require the presence
of a strong compact component with a flux density of $\sim10$~mJy
offset by $\sim3$~mas from the stellar disc center.  A gaussian
component of $\sim4.7$~mas ($\sim0.4$~AU) produces a good fit, though
cannot rule out a cluster of delta functions. We have thus detected
the presence of a significant hotspot, or a compact cluster of spots,
on the Mira A stellar disc at $\sim1.3$~mm wavelength. We have
investigated if the same spot could be detected in the Band 3
observations at $\sim3$~mm.  Fits including a gaussian component up to
$\sim2.5$~mJy produced equally good results, but the worse angular
resolution did not allow us to confidently distinguish between a fit
of a stellar disc plus compact component or a stellar disc alone.
 
The flux density measurements allow us to determine the
brightness temperature. With a size of $\sim4.7$~mas,
we find $T_b\sim10000$~K, above the brightness temperature of $\sim2500$~K
measured for the stellar disc in Band 6. The upper limit in Band 3,
assuming a 4.7~mas area, gives $T_b<17500$~K. Bright hotspots at millimeter wavelenghts could be
caused by shock heating due to pulsations or convection. However, the
high brightness temperature of the hotspots at
$\sim1.3$~R$_*$ is more readily explained by magnetic activity as seen
on our Sun. Similar magnetic flares were also
suggested to be the cause of the soft X-ray outburst observed on Mira
A in December 2003 \citep{karoetal05} but this is the first direct
detection of such magnetic activity of an AGB star in the long
wavelength regime. The observed hotspot
could be related with the strongly polarized elongated SiO
maser ejections that appear aligned with a radial magnetic field
\citep{Cotton06}. \citet{karoetal05} crudely estimated the lifetime of
such magnetic flares to be of order several weeks. We notice a
significant difference of almost 15\% within 2~days. This would be
consistent with the lifetime estimate of a few weeks, but could also
indicate significant short term variability.

\subsection{Mira B}

The accreting companion Mira B has a time-varying accretion-driven
wind ($\dot{M} \sim 5\times10^{-13}-10^{-11}\,M_{\odot}$\,yr$^{-1},
v_{\infty} \sim 250-450$\,km\,s$^{-1}$, \citealt{woodkaro06}) and
recently ALMA showed the impact of this fast and tenuous wind on the
much slower and denser outflow ($\dot{M} \sim
10^{-7}\,M_{\odot}$\,yr$^{-1}, v_{\infty} \sim 10$\,km\,s$^{-1}$) of
Mira A \citep{ramsetal14}. Even if fully ionized, the expected flux
density of Mira B's wind would be at least 4 orders of magnitude
smaller than that we observe at 89 and 229\,GHz, and would also be
optically thin (i.e., $S_\nu \propto \nu ^{-0.1}$). We therefore rule
out any significant contribution from Mira B's wind at ALMA
frequencies. The steep spectral index of Mira B ($\alpha \sim
1.3-1.7$) also rules out any significant non-thermal component,
while the symmetric nature of the resolved emission makes it unlikely
that the emission is being produced from a collision of the two
winds. Our derived diameter for Mira B is $\sim2.4\,$AU which is 2
orders of magnitude greater than the size of the accretion disk
detected by \cite{reimcass85} and so we do not directly detect the
accretion disk at ALMA frequencies.

\cite{mattkaro06} treated the centimeter emission from Mira B to
emanate from an ionized cavity in Mira A's wind, formed by the UV
radiation field of the hot accretion disk. They approximated a
diameter of $1-5$\,AU for Mira B, which is consistent with our
findings. However, following \cite{taylor_1984} and assuming a gas
temperatures of $T=10^4\,$K, the spectral energy distribution for such
an ionized cavity should have turned over to a spectral index of
$\alpha = -0.1$ well below our ALMA frequencies. Instead we find that
$\alpha = 1.3\pm0.2$ at 89\,GHz, which is consistent with the $\alpha
= 1.18\pm 0.28$ value of \cite{mattkaro06} at lower frequencies. The
absence of a turnover in the spectral index up to 89\,GHz implies that
the gas density at the location of Mira B is at the very least an
order of magnitude greater than the gas density of Mira A's wind at
that location. It is therefore likely that we are detecting a
gravitationally bound partially ionized gas at centimeter and
(sub-)millimeter wavelengths and not an ionized cavity of Mira A's
wind. We also find that the spectral index steepens to $\alpha =
1.72\pm0.11$ between 89 and 229\,GHz, which could be due to additional
dust emission at a few tens of K.

\section{Conclusions}

With the longest ALMA baselines we have, for the first time, been able
to resolve both the extended atmosphere of Mira A and the partially
ionized region around Mira B at millimeter wavelengths. The partially
ionized region of $\sim2.4$~AU around Mira B requires a density at
least an order of magnitude higher than expected from the Mira A
wind. This supports the suggestion that the region is made up from
gravitationally captured material from the AGB wind that will be
accreted on Mira B through its accretion disk. The elongation and
wavelength dependent size of Mira A point to strong changes of opacity
source throughout the extended envelope and possible non-radial
pulsations. The brightness temperature at 3~mm is significantly higher
than predicted in the radio photosphere model from
\citet{reidment97}. At 1.3~mm wavelength we also find a hotspot with
an area of $\sim8\%$ of the stellar disc. The hotspot has
$T_b\sim10000$~K. This indicates an origin likely connected to
magnetic activity and potentially related to the process responsible
for the previously observed X-ray outburst and other recorded
transient phenomena. The stellar activity could also explain the hot
layers of the atmosphere probed by the 3~mm observations.

\begin{acknowledgements} 
  This paper makes use of the following ALMA data:
  ADS/JAO.ALMA\#2011.0.00014.SV . ALMA is a partnership of ESO
  (representing its member states), NSF (USA) and NINS (Japan),
  together with NRC (Canada) and NSC and ASIAA (Taiwan), and KASI
  (Republic of Korea), in cooperation with the Republic of Chile. The
  Joint ALMA Observatory is operated by ESO, AUI/NRAO and NAOJ. This
  research is supported by Marie Curie Career Integration Grant
  321691, the Swedish Research Council (VR) and the European Research
  Council through ERC consolidator grant 614264.
\end{acknowledgements}

\newpage

\begin{appendix}

%

\section{Flux density calibrators}
\label{flux}

Several quasars were used for flux density calibration at the different epochs
and bands. In Table.\ref{tableflux} we present the calibrators, and their
adopted flux densities and spectral indices from the ALMA calibrator catalog.

\begin{table}
\caption{Calibrator fluxes}
\begin{center}
\begin{tabular}{lcccc}
\hline \hline
epoch & calibrator & reference freq. & flux density & spectral index \\
      &  & [GHz] & [Jy] & \\
\hline \hline
17 Oct 2014  & J0334-4008 & 86.23 & 1.66 & -0.71 \\  
25 Oct 2014  & J2258-2758 & 86.23 & 1.21 & -0.73 \\
29 Oct 2014  & J0334-4008 & 229.55 & 0.83 & -0.70 \\
01 Nov 2014 &  J0238+1636 & 229.55 & 1.39 & -0.20 \\
\end{tabular}
\end{center}
\label{tableflux}
\end{table}%

\section{Relative astrometry}
\label{relastrom}

The binary pair Mira AB is fully resolved in the observations, with a
representative image at the two frequencies presented in
Fig.~\ref{fig:Mira2} (left) and Fig.~\ref{fig:Mira} (left). Using $uv$-fitting we determine the
separation and position angle with milliarcsecond accuracy as
indicated in Table~\ref{astrom}. The average separation of
0.472\arcsec corresponds to $43.4$~AU at a distance of $92$~pc. These
values fit well with the predictions made by \citet{Prieur02},
although the authors note that their derived binary orbit is still
poorly constrained. It is interesting to note that the data already
show a (not yet significant) sign of a decrease in separation $R$ over
two weeks of observing at a rate of
$\sim121\pm71~\mu$as~day$^{-1}$. Futher epochs of observations with
  ALMA will be thus be able to constrain the binary orbit to high
  precision.

\begin{table}[hb]
\caption{Relative astrometry}
\begin{center}
\begin{tabular}{lcc}
\hline \hline
epoch & separation & position angle \\
      & [\arcsec]  & [$^{\circ}$] \\
\hline \hline
17 Oct 2014  & 0.4722$\pm$0.0005 & 98.79$\pm$0.07\\  
25 Oct 2014  & 0.4721$\pm$0.0005 & 98.59$\pm$0.06\\ 
29 Oct 2014  & 0.4719$\pm$0.0002 & 98.64$\pm$0.02\\  
01 Nov 2014 & 0.4709$\pm$0.0001 & 98.53$\pm$0.02\\
\end{tabular}
\end{center}
\label{astrom}
\end{table}%

\end{appendix}

\end{document}